\documentclass[letterpaper]{article}
\usepackage[T1]{fontenc}
\usepackage{comment}
\usepackage{geometry}
\usepackage{setspace}

\usepackage[articletitle = true, chaptertitle = true]{achemso} 
\usepackage{graphicx}
\usepackage{float}
\newfloat{scheme}{htbp}{los}
\floatname{scheme}{Scheme}
\floatname{chart}{Chart}
\newfloat{graph}{htbp}{loh}
\usepackage{lineno}

\usepackage{chemformula} 
\usepackage[version = 4]{mhchem} 

\usepackage[hidelinks]{hyperref}

\title{Surface-Sensitive Mapping of Anisotropic Phonon Cascades in T$_d$-WTe$_2$}

\usepackage{authblk}

\author[1,2]{Alp Akbiyik}
\author[1,2]{Felix Kurtz}
\author[1,2]{Nandana Veena Uday}
\author[1,2]{Sergey V. Yalunin}
\author[1,2]{Claus Ropers}
\author[1,2,*]{Hannes Böckmann}

\affil[1]{Department of Ultrafast Dynamics, Max Planck Institute for Multidisciplinary Sciences, 37077 Göttingen, Germany}
\affil[2]{IV. Physical Institute, Solids and Nanostructures, University of Göttingen, 37077 Göttingen, Germany}

\date{*Email: hannes.boeckmann-clemens@mpinat.mpg.de}

\begin{document}

\maketitle
\baselineskip24pt
\begin{abstract}
\baselineskip24pt
Understanding how energy flows from photoexcited carriers into the lattice is essential for describing nonequilibrium phenomena in low-symmetry quantum materials. Here, we use ultrafast low-energy electron diffraction and diffuse scattering to probe momentum-resolved phonon dynamics at the surface of T$_d$-WTe$_2$, a strongly anisotropic semimetal. Following optical excitation, the Debye--Waller suppression of Bragg peaks exhibits a biexponential increase of the mean-squared atomic displacement, indicating sequential lattice relaxation. Analysis of the diffuse background reveals a preferential intensity build-up parallel to the tungsten-chain axis in the material, attributed to anisotropic electron--phonon coupling during electronic cooling which precedes anharmonic phonon--phonon scattering and subsequent thermalization across the surface Brillouin zone. The results identify a hierarchical relaxation pathway in which energy is first deposited into selected finite-momentum phonons before spreading through the broader lattice bath. Our work highlights the importance of momentum-resolved diffuse scattering for disentangling electron--phonon and phonon--phonon relaxation in anisotropic topological semimetals.
\end{abstract}
\section*{Keywords}

T$_d$-WTe$_2$; ultrafast electron diffraction; diffuse scattering; phonon dynamics; electron--phonon coupling; nonequilibrium materials; topological semimetals


\section{Introduction}

\begin{figure*}[t]
    \includegraphics[width=0.5\textwidth]{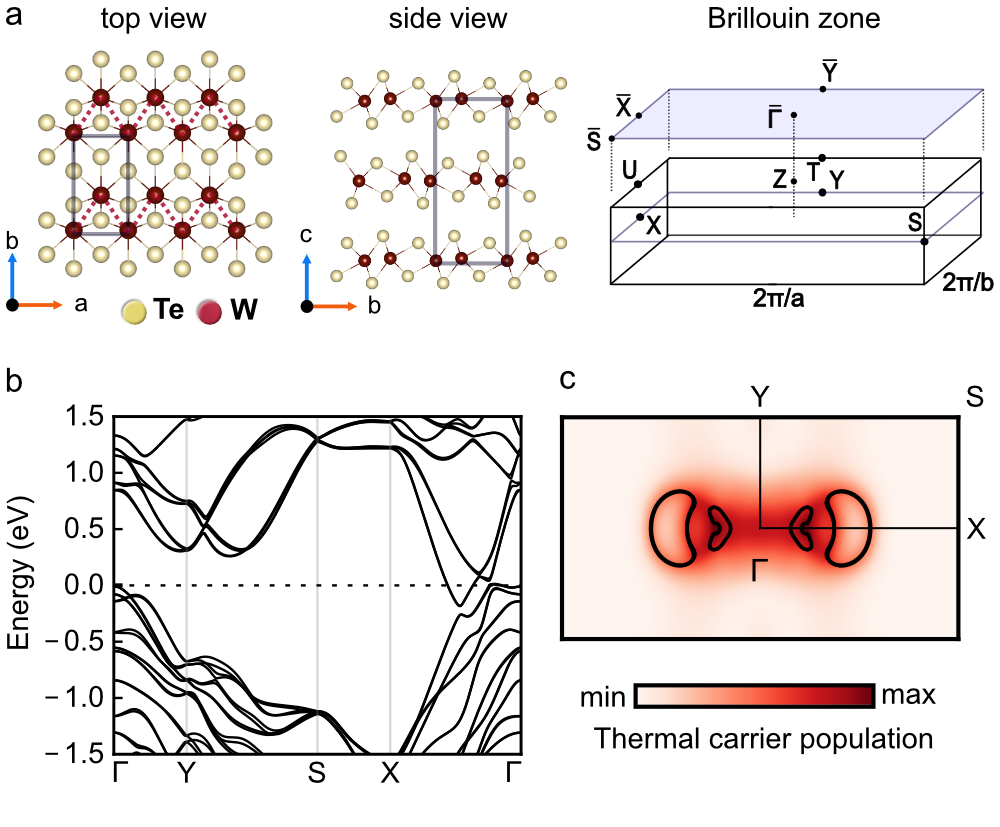}
\caption{(a) Top and side view of the T$_{d}$-WTe$_{2}$ atomic structure and the corresponding Brillouin zone schematic \cite{vesta}. The purple area represents the surface Brillouin zone and corresponding high-symmetry points. (b) Electronic band structure of bulk T$_{d}$-WTe$_{2}$, obtained from density functional theory calculations. (c) Momentum distribution of thermalized charge carriers, following a Fermi-Dirac distribution. Black lines indicate the Fermi surface and the location of electron and hole pockets.} 
\label{fig:structure}
\end{figure*}

Layered transition-metal ditellurides provide a versatile platform for studying how anisotropic lattice dynamics shape nonequilibrium electronic properties. In its orthorhombic T$_{d}$ phase, WTe$_{2}$ combines broken inversion symmetry, strong spin-orbit coupling, small charge-compensated electron and hole pockets, and pronounced in-plane structural anisotropy. These features underlie its giant magnetoresistance \cite{Ali2014, Pletikosic2014} and the classification as a type-II Weyl semimetal candidate \cite{Soluyanov2015}. At the same time, the close interplay between electronic and lattice degrees of freedom makes the electronic structure highly sensitive to small atomic displacements. In particular, the W atoms arrange in chains along the crystallographic a-axis, defining a preferred in-plane direction and giving rise to strongly anisotropic electronic and vibrational properties (Fig. \ref{fig:structure}a) \cite{Mar1992WTe2, Kim_2016}. Understanding how photoexcited carriers transfer energy to this anisotropic lattice is therefore essential for describing the transient properties of WTe$_{2}$ and related low-symmetry semimetals. Time-resolved studies have shown that optical excitation drives rapid electronic redistribution and coherent lattice motion \cite{Sie2019, Hein2020}. Moreover, selected strongly-coupled lattice modes can modify the symmetry and topology of the band structure, demonstrating the capacity for active control over the material’s electronic state \cite{Sie2019, Ji2021}. Because of the intrinsic electronic and structural anisotropy of T$_{d}$-WTe$_{2}$, however, the initial phonon population is expected to depend significantly on phonon momentum, branch character, and crystallographic direction. \\

Momentum-resolved electron scattering provides a direct route to disentangle these relaxation pathways. In particular, diffuse scattering from incoherent finite-momentum phonons can distinguish populations mediated by electron-phonon (el-ph) scattering from those that build up through anharmonic phonon-phonon (ph-ph) relaxation \cite{Chase2016, Waldecker, Stern2018, maldonado2020, Otto2021, Cheng2024}. This distinction is especially important in T$_{d}$-WTe$_{2}$, where mode- and momentum-selective el-ph coupling has been identified as a key ingredient of the nonequilibrium response \cite{Hein2020}. While previous ultrafast diffraction studies have revealed coherent zone-center lattice motion through Bragg-peak modulations \cite{Sie2019, Ji2021}, such coherent phonons represent only part of the lattice response. The incoherent finite-momentum phonons, which dominate thermalization and energy flow, remain less explored. \\

Here, we use ultrafast low-energy electron diffuse scattering (ULEEDS) \cite{Kurtz2024} to track momentum-resolved nonthermal phonon populations in photoexcited T$_{d}$-WTe$_{2}$. We observe a sequential relaxation pathway with pronounced in-plane anisotropy. The initial response is strongly directional and associated with preferential el-ph coupling along the W chain direction. Anharmonic redistribution of population across the Brillouin zone is followed by a gradual buildup of zone-center acoustic modes, mediated by ph-ph scattering on a 30--100 ps timescale. These results show that energy relaxation in T$_{d}$-WTe$_{2}$ proceeds through a hierarchy of momentum-selective processes rather than through homogeneous lattice thermalization, highlighting the central role of anisotropic el-ph interactions in low-symmetry topological semimetals. \\

The crystal structure of T$_{d}$-WTe$_{2}$ consists of distorted WTe$_{2}$ layers stacked along the c-axis in a non-centrosymmetric arrangement (Fig.1a). The low-symmetry T$_{d}$ polymorph is stabilized below the first-order structural transition near T$_c$=565~K \cite{Mar1992WTe2, Tao2020} and exhibits a trilayer-like stacking with broken inversion symmetry along the out-of-plane direction. Within each layer, covalent W-W bonding along the a-axis produces a quasi-one-dimensional structural motif and a pronounced in-plane anisotropy \cite{TangZhou2019, Song2016}. Correspondingly, the low-energy electronic structure consists of small, charge-compensated electron and hole pockets near the Fermi level (Fig. 1b), forming an elongated semimetallic Fermi surface that reflects the W-chain direction (Fig. 1c) (see Methods: Density-functional-theory calculations). This anisotropic electronic structure provides a natural basis for directional el-ph scattering and momentum-selective phonon generation. \\

Ultrafast optical excitation, spanning from the THz to visible spectral regime, has been shown to launch coherent lattice modes associated with interlayer sliding and structural distortions \cite{Hein2020, Aoki2022, He2016, Dai2015}. These modes can transiently drive the system towards a topologically trivial, high-symmetry configuration, and modulate diffraction intensities and optical properties \cite{Ji2021, Drueke2021, Soranzio2022}. Momentum-resolved electronic and structural probes thereby link coherent Bragg peak modulation to band-selective el-ph interactions near the electron and hole pockets \cite{Sie2019, Ji2021, Hein2020}, yet coherent modes represent only part of the lattice response. Incoherent finite-momentum phonon populations encode where energy is deposited by hot carriers and how it subsequently spreads. Direct probing of these populations can be used to map out thermalization processes in momentum space.

\section{Results and Discussion}

For this purpose, we employ ultrafast low-energy electron diffraction (ULEED) (Fig. 2a) \cite{Vogelgesang2018, storeck2020} (see Methods: Ultrafast LEED Measurements). The technique combines the high surface sensitivity of photoemitted low-energy electrons with femtosecond optical excitation in a backscattering geometry, enabling direct access to surface structural dynamics and non-equilibrium phase evolution \cite{storeck2020, Horstmann2020, boeckmann2022, Boeckmann2025}. More recently, momentum-resolved analysis of the diffuse background has been used to probe non-thermal phonon dynamics \cite{Kurtz2024}. The transient change in diffuse-scattering signal arises from inelastic electron scattering by phonons. In this process, probe electrons exchange momentum with lattice vibrations and get deflected from ideal Bragg spot positions by the phonon wave vector, which manifests in a transient decrease of Bragg spot intensity and corresponding increase in diffuse background (Fig. 2b) \cite{Waldecker}. As the scattering vector in ULEED is oriented predominantly perpendicular to the surface and varies only weakly across the detector, the measurement is particularly sensitive to phonon modes with out-of-plane displacement components \cite{Kurtz2024}. In addition, low-frequency modes contribute most strongly to the diffuse intensity due to the inverse frequency scaling of the phonon structure factor \cite{Xu2005}. The resulting time-dependent diffuse intensity maps therefore provide access to structure-factor-weighted transient phonon populations across the surface Brillouin zone. Consequently, diffuse scattering in ULEED is particularly sensitive to dynamics in the ZA branch of the phonon band structure (Fig. 2c).\\
As a key observation, we find that the phonon population build-up in T$_{d}$-WTe$_{2}$ depends not only on the magnitude of the momentum transfer, but also on the crystallographic direction. An iso-momentum cut through the surface Brillouin zone reveals accelerated dynamics along the W-chain direction with respect to perpendicular direction which warrants further investigation (Fig. 2d).\\

\begin{figure*}
\includegraphics[width=\textwidth]{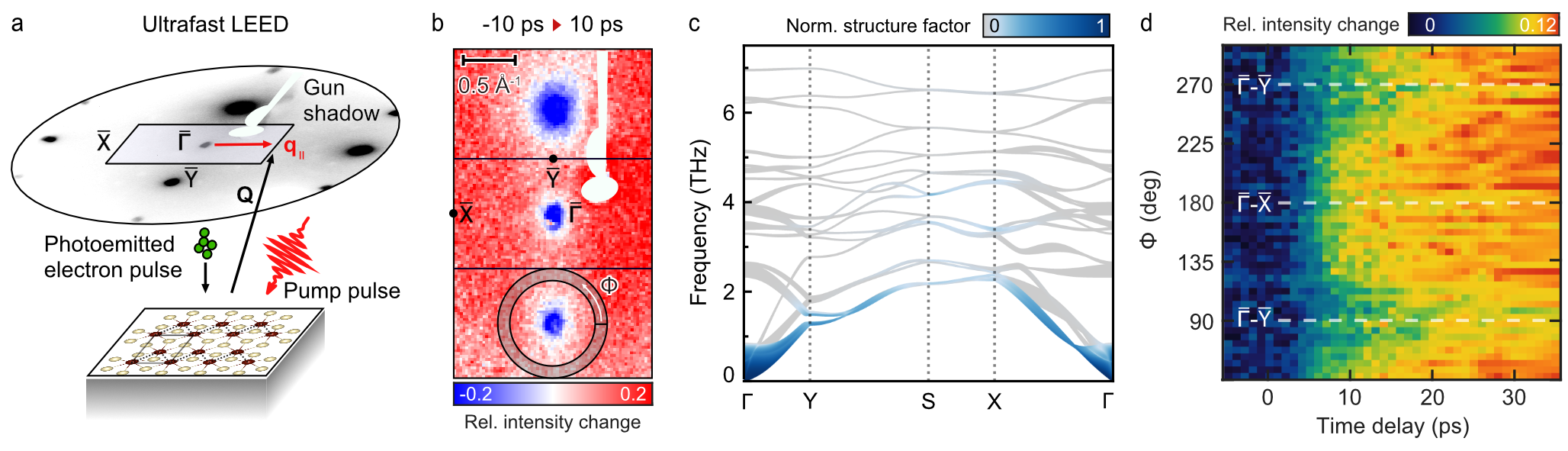}
\caption{(a) Schematic of the ULEED experimental setup. Back-scattered low-energy electrons along the scattering vector $\mathbf{Q}$ reveal Bragg spot suppression and a diffuse background increase at phonon momenta $\mathbf{q}_{||}$. (b) Difference image of pump-induced changes in diffraction intensities. (c) Phonon dispersion for T$_{d}$-WTe$_{2}$, color-coded by the normalized one-phonon structure factor, emphasizing the experimental sensitivity to low-energy acoustic modes. (d) Time-dependent change in diffraction intensity at constant $||\mathbf{q}_{||}||$ as function of azimuthal angle $\Phi$ around the Bragg spot. The analyzed intensities are indicated in (b) and averaged across surface Brillouin zones.}
\label{fig:msd}
\end{figure*}

We first quantify the transient phonon excitation through the Debye-Waller suppression of the Bragg peaks (Fig. 3a) \cite{VanHove1986LEED}. By averaging the pump-induced intensity reduction over the diffraction peaks, we extract the time-dependent change in mean-squared displacement (MSD) as an incoherent, structure-factor-weighted measure of the lattice response (see Supporting Information: MSD Change and Diffuse Background Dynamics). For T$_{d}$-WTe$_{2}$, the MSD exhibits a biexponential increase, revealing sequential population dynamics of the modes contributing to the Debye-Waller response (Fig. 3b). An initial rapid rise within the first few picoseconds is followed by a slower increase on a 30~ps timescale, which accelerates at increasing sample temperature (see Supporting Information: Fig. S1). The clear temporal separation of these two components suggests that the early response is governed predominantly by el-ph scattering during electronic cooling, whereas the delayed component reflects subsequent ph-ph redistribution and the gradual buildup of lower-energy lattice populations. This assignment is further supported by the fluence dependence of the extracted time constants (Fig. 3c). Electron-phonon energy transfer rates are typically found to depend only weakly on excitation density, so that the fast rise time remains approximately fluence independent \cite{mueller2014}. Anharmonic scattering among lattice modes nonlinearly depends on occupation such that increasing phonon populations enhance the available scattering phase space and accelerate the redistribution process \cite{Fugallo2013}. Accordingly, we find a linear scaling of the scattering rate with increasing excitation density, thus supporting its attribution to ph-ph relaxation. \\
\begin{figure*}
\includegraphics[width=\textwidth]{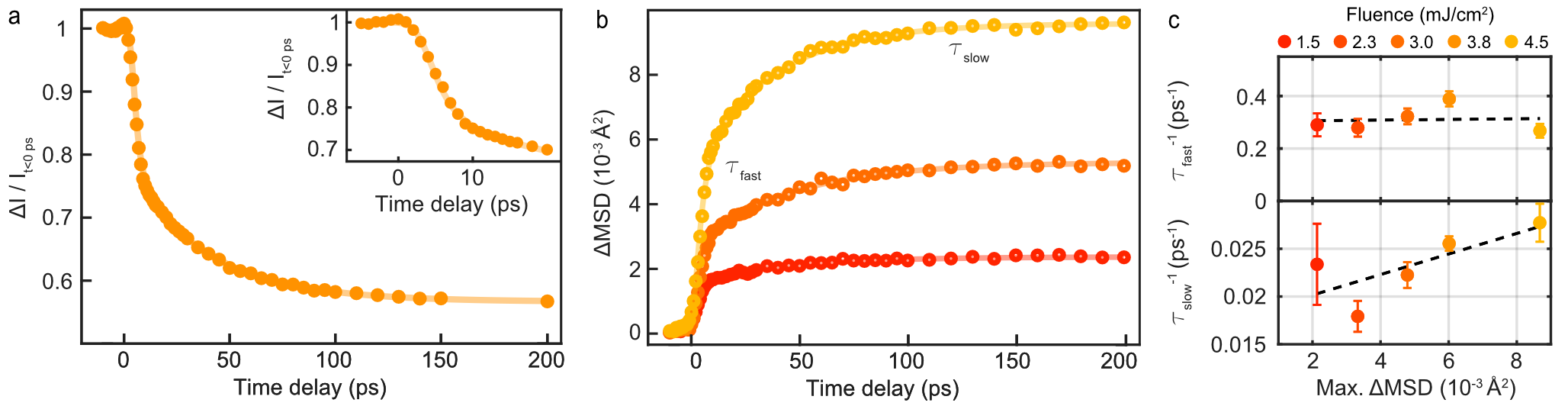}
\caption{(a) Pump-induced suppression of Bragg peak intensity due to the Debye-Waller effect. (b) Transient change in out-of-plane mean-squared displacement (MSD) for selected absorbed fluences. Biexponential fits capture the dynamics, indicating a cascaded rise in distinct phonon populations. (c) Fluence-dependent time constants for the fast and slow component of the MSD increase, obtained from the fit model.}
\label{fig:diff}
\end{figure*}

Having separated the different contributions to the MSD rise, we next determine how the corresponding phonon populations are distributed in momentum space by analyzing the diffuse background (see Supporting Information: MSD Change and Diffuse Background Dynamics). Since the total number of scattered electrons is approximately conserved, the pump-induced Debye-Waller suppression of the Bragg peaks is accompanied by an increase in diffuse intensity, distributed across the surface Brillouin zone. We focus the analysis on the sequential population dynamics along the principal in-plane crystallographic directions, \textbf{a} and \textbf{b}, corresponding to the $\bar{\Gamma}-\bar{X}$ and $\bar{\Gamma}-\bar{Y}$ directions, respectively. To quantify these dynamics, we integrate the diffuse intensity within selected regions along the zone boundaries and track the phonon thermalization to lowest-frequency modes close to the zone center (Fig. 4a).  The resulting time traces reveal an anisotropic redistribution of vibrational energy through the lattice. Around the $\bar{X}$-point, we observe a prompt intensity increase on a 3.5 ps timescale, which we associate with the fast component of the MSD rise. Its pronounced momentum selectivity, together with its weak dependence on excitation fluence, identifies anisotropic el-ph coupling as the primary source of this early phonon population. This assignment is consistent with the electronic structure of T$_{d}$-WTe$_{2}$, where elongated electron and hole pockets and relevant band extrema are aligned predominantly along the $\bar{\Gamma}-\bar{X}$ direction, providing substantial phase space for carrier relaxation by finite-momentum phonons \cite{Pletikosic2014, Hein2020, Ji2021, Jiang2015}. In contrast, the diffuse intensity around the $\bar{Y}$-point exhibits a delayed buildup that evolves on a 10 ps timescale, preceding a subsequent increase of intensity near the zone center that saturates only after more than 100 ps. These slower components mirror the delayed contribution to the MSD change and indicate that phonon populations outside the initially coupled $\bar{\Gamma}-\bar{X}$ regions are generated primarily through anharmonic ph-ph scattering. A pixelwise extraction of the corresponding time constants yields a detailed map that reflects the cascaded energy flow within the Brillouin zone (Fig. 4b). The momentum-selective build-up of phonon populations manifests as systematically larger rise times along the $\bar{\Gamma}-\bar{Y}$ direction than along $\bar{\Gamma}-\bar{X}$, as extracted from selected cuts through the Brillouin zone (Fig. 4c). This crystallographic dependence contrasts with a previous ULEEDS study on TiSe$_2$~\cite{Kurtz2024}, and directly reflects the structural and electronic anisotropy of T$_{d}$-WTe$_{2}$. \\

\begin{figure*}
\includegraphics[width=\textwidth]{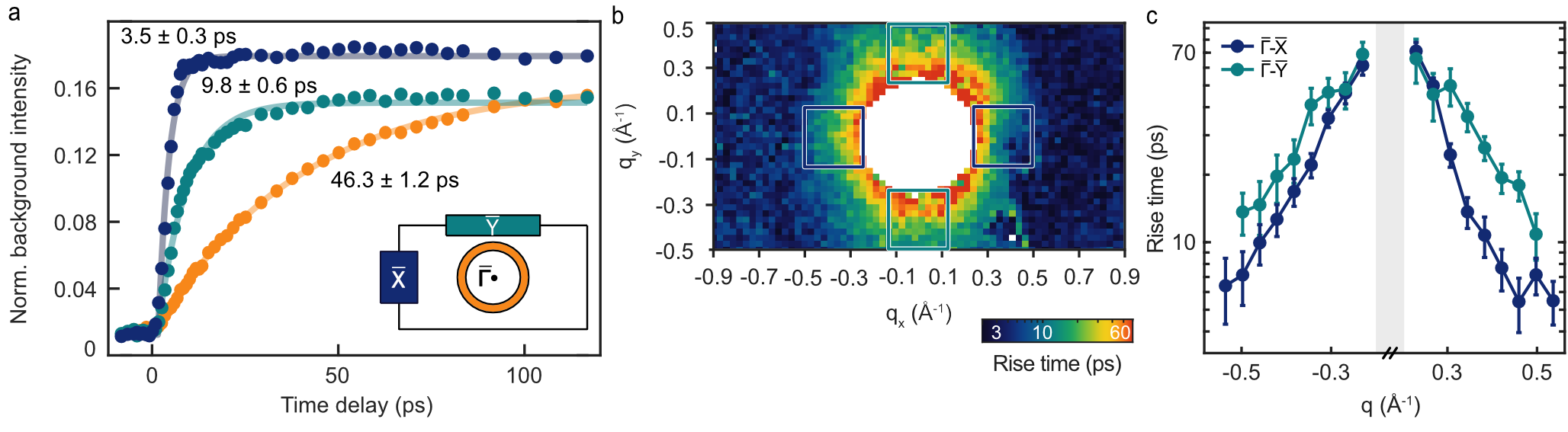}
\caption{(a) Transient change in diffuse background, evaluated at the Brillouin zone $\bar{X}$- and $\bar{Y}$-boundaries (blue and turquoise) and close to the zone center (orange) (Fluence: 3.8 mJ/cm$^{2}$). Time constants are obtained from single-exponential fits. (b) Phonon rise time map, deduced from a spatially-binned pixelwise analysis. Bragg peak dynamics are excluded manually within an area of 4.5$\sigma$ around the spot, where $\sigma$ denotes the spot width. (c) Rise times, evaluated within momentum cuts along $\bar{X}$- and $\bar{Y}$-directions, as indicated in (b).}
\label{fig:bgfit}
\end{figure*}
In summary, the excitation of coherent vibrational modes in T$_{d}$-WTe$_{2}$ is directly linked to low-energy electronic states, connected primarily along the $\bar{\Gamma}-\bar{X}$ direction, which imposes momentum selectivity in the phonon emission (Fig. 1c) \cite{Hein2020}. As these collective oscillations dephase or decay within the first few picoseconds, their energy is transferred into incoherent finite-q phonons with anisotropic distribution. However, Bragg-peak dynamics are primarily sensitive to coherent zone-center displacements and average structural changes while ULEEDS measurements resolve subsequent thermalization processes. In this way, diffuse scattering tracks the redistribution of vibrational energy into the broader phonon bath. The delayed population growth along $\bar{\Gamma}-\bar{Y}$ signifies this equilibration while the increase of zone-center acoustic intensity reflects the gradual transfer of energy into low-frequency modes (Fig. 5). 

\begin{figure*}
\includegraphics[width=0.5\textwidth]{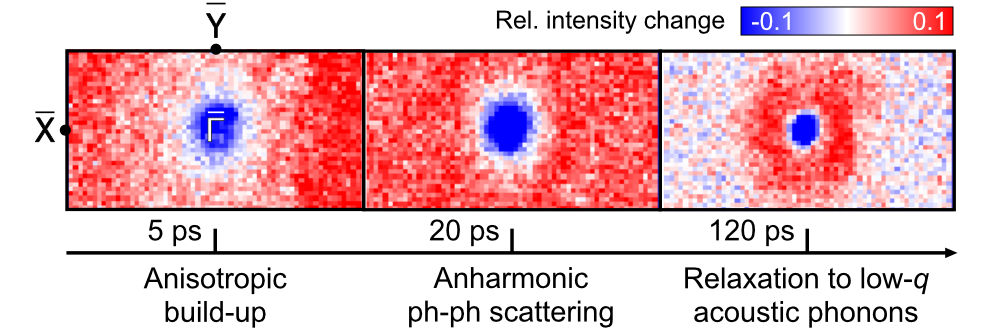}
\caption{Diffuse background evolution within the Brillouin zone between different time steps. An el-ph scattering mediated anisotropic build-up in $\bar{X}$-direction is followed by anharmonic redistribution across the Brillouin zone. The delayed build-up at the zone center reflects scattering towards low-q acoustic phonons.}
\label{fig:summary}
\end{figure*}

\section{Conclusion}

Previous ultrafast studies on T$_{d}$-WTe$_{2}$ have emphasized coherent structural and electronic motion and its impact on symmetry and topology \cite{Sie2019, Ji2021, He2016, Dai2015}. Time-resolved photoemission and electron diffraction results showcase the coherent excitation of zone-center optical modes by displacive and impulsive forces, most prominently a low-frequency interlayer sliding mode \cite{Hein2020, Sie2019, Ji2021}. As this mode involves predominantly in-plane relative displacements of adjacent layers, we attribute the missing signature of this modulation in our data to the surface confinement of low-energy electron elastic scattering and the out-of-plane sensitivity of backscattering geometries. Instead, our study expands upon prior works by directly probing the incoherent finite-momentum phonons that govern lattice thermalization. The sequential dynamics observed here elucidate the role of momentum-selective el-ph coupling for subsequent lattice equilibration and are consistent with previous electron diffraction studies on anisotropic materials such as black phosphorus \cite{Zahn2020, Seiler2021}. More generally, ultrafast low-energy electron diffuse scattering provides a surface-sensitive perspective to transmission ultrafast electron diffraction and x-ray diffuse scattering. Its sensitivity to low-frequency and out-of-plane polarized modes makes it especially suited for layered materials, surfaces and heterostructures, where interlayer coupling, surface phonons, and symmetry-breaking distortions can differ substantially from the bulk  \cite{Theuss2024, Tinnemann2019, Zhu2018, Kurtz2025}. 

\section{Methods}

\subsection{Density Functional Theory Calculations}
First-principles calculations were performed with density functional theory (DFT) using the Quantum ESPRESSO suite of codes~\cite{Giannozzi2017} including phonons from perturbation theory~\cite{Baroni2001}. We employed the generalized gradient approximation (GGA) with the Perdew-Burke-Ernzerhof exchange-correlation functional (PBE), optimized norm-conserving Vanderbilt pseudopotentials~\cite{Hamann2013, vanSetten2018}. The long-range van der Waals interactions were included via the DFT-D3 correction scheme~\cite{dft}. All calculations were performed using a fully optimized crystal structure with lattice parameters very close to their experimental values~\cite{Mar1992}. We used a plane wave kinetic-energy cutoff value of 80 Ry, and the electronic and vibrational Brillouin zones were sampled using a uniform grid of 18$\times$10$\times$4 and 3$\times$2$\times$2 points, respectively. For the Debye-Waller factor calculation, we interpolated the phonon frequencies and polarization vectors on a finer grid of 60$\times$40$\times$10 points. Details on the phonon structure factor calculation can be found in the Supporting Information.

\subsection{Ultrafast LEED Measurements}
ULEED measurements are carried out in an optical pump and electron probe configuration where electrons are photoexcited by 40 fs pulses at 400 nm out of the micron-scale electron gun, introduced in earlier works \cite{storeck2017, Vogelgesang2018}. The instrument is characterized by a transfer width of ~9.3 nm while the time resolution is limited by the electron beam spread along the gun-sample distance and amounts to about 3 ps in the presented measurements. The electron energy is 90 eV and the beam spot size on the sample is about 10 µm. Backreflected electrons from the sample are collected on the detection unit consisting of microchannel plates (MCP), phosphor screen, and CMOS camera. The pump beam has a central wavelength of 1030 nm  (200 fs pulsewidth at 100 kHz repetition rate). The laser beam is incident on the sample at about 45$^\circ$, focused down to 200 $\mu$m diameter. Further details on the pump fluence calculation can be found in the Supporting Information.

Bulk WTe$_2$ crystals are obtained commercially from HQ Graphene \cite{HQGraphene} and the surface is prepared via tape exfoliation prior to introduction into the UHV chamber. Samples are mildly heated up to 500 K while characterizing the surface in front of a commercial LEED (SPECS) before  ULEED measurements. The characterization and measurements are carried out at a base pressure of $2\times10^{-10}$ mbar. Measurements are taken at 30 K base temperature.


\section{Acknowledgments}
This work is funded by the European Research Council (ERC Advanced Grant “ULEEM,” ID: 101055435). The authors thank J.G Horstmann for the fruitful discussions.

\section*{Supporting information}

The following files are available free of charge.
\begin{itemize}
  \item Supporting information : MSD Change and Diffuse Background Dynamics; Pump Fluence Calculation; Phonon Structure Factor Calculation
\end{itemize}

\section*{Data Availability}
The data presented in the manuscript are available in Edmond-the Open Research Data Repository of the Max Planck Society via \url{https://doi.org/10.17617/3.JONMZP} \cite{data_AlpAkbiyik}.

\clearpage

\section*{Table of Contents Graphic}

\begin{figure}[H]
    \centering
    \includegraphics[width=70mm]{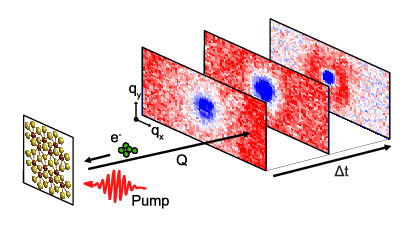} 
\end{figure}

\clearpage

\bibliography{references}

\end{document}